\documentclass[twocolumn,english,aps,prb,superscriptaddress]{revtex4-1}
\usepackage{grffile}
\usepackage[latin9]{inputenc}
\usepackage{color}
\usepackage{babel}
\usepackage{graphicx}
\usepackage{epstopdf}
\usepackage{cancel}
\PassOptionsToPackage{normalem}{ulem}
\usepackage{ulem}
\usepackage[unicode=true, bookmarks=false, breaklinks=false,pdfborder={0 0 1},colorlinks=false] {hyperref}
\usepackage{breakurl}

\begin{document}

\title{Growth, characterization and Chern insulator state in MnBi$_2$Te$_4$ via the chemical vapor transport method}

\author{Chaowei Hu}
\affiliation{Department of Physics and Astronomy and California NanoSystems Institute, University of California, Los Angeles, CA 90095, USA}
\author{Anyuan Gao}
\affiliation{Department of Chemistry and Chemical Biology, Harvard University, Cambridge, MA, USA}
\author{Bryan Stephen Berggren}
\affiliation{Department of Physics and Center for Experiments on Quantum Materials, University of Colorado, Boulder, CO 80309, USA}
\author{Hong Li}
\affiliation{Department of Physics, Boston College, Chestnut Hill, MA 02467, USA}
\author{Rafa\l{} Kurleto}
\affiliation{Department of Physics and Center for Experiments on Quantum Materials, University of Colorado, Boulder, CO 80309, USA}
\author{Dushyant Narayan}
\affiliation{Department of Physics and Center for Experiments on Quantum Materials, University of Colorado, Boulder, CO 80309, USA}
\author{Ilija Zeljkovic}
\affiliation{Department of Physics, Boston College, Chestnut Hill, MA 02467, USA}
\author{Dan Dessau}
\affiliation{Department of Physics and Center for Experiments on Quantum Materials, University of Colorado, Boulder, CO 80309, USA}
\author{Suyang Xu}
\affiliation{Department of Chemistry and Chemical Biology, Harvard University, Cambridge, MA, USA}
\author{Ni Ni}
\email{Corresponding author: nini@physics.ucla.edu}
\affiliation{Department of Physics and Astronomy and California NanoSystems Institute, University of California, Los Angeles, CA 90095, USA}

\begin{abstract}

As the first intrinsic antiferromagnetic topological insulator, MnBi$_2$Te$_4$ has provided a platform to investigate the interplay of band topology and magnetism as well as the emergent phenomena arising from such an interplay. Here we report the chemical-vapor-transport (CVT) growth and characterization of MnBi$_2$Te$_4$, as well as the observation of the field-induced quantized Hall conductance in 6-layer devices. Through comparative studies between our CVT-grown and flux-grown MnBi$_2$Te$_4$ via magnetic, transport, scanning tunneling microscopy, and angle-resolved photoemission spectroscopy measurements, we find that CVT-grown MnBi$_2$Te$_4$ is marked with higher  Mn occupancy on the Mn site, slightly higher Mn$_{\rm{Bi}}$ antisites, smaller carrier concentration and a Fermi level closer to the Dirac point. Furthermore, a 6-layer device made from the CVT-grown sample shows by far the highest mobility of 2500 cm$^2$V$\cdot$s in MnBi$_2$Te$_4$ devices with the quantized Hall conductance appearing at 1.8 K and 8 T. Our study provides a new route to obtain high-quality single crystals of MnBi$_2$Te$_4$ that are promising to make superior devices and realize emergent phenomena, such as the layer Hall effect and quantized anomalous hall effect, etc. \end{abstract}

\date{\today}
\maketitle
\section{Introduction}
Magnetic topological insulators (MTI) are a rising family of quantum materials that host both nontrivial band topology and magnetic order, providing a fruitful platform to realize emergent quantum phenomena such as the quantum anomalous Hall effect (QAHE), axion insulator state, quantum magnetoelectric effect, etc. \cite{he2018topological,liu2016quantum,wang2015quantized} Recently, MnBi$_2$Te$_4$ was proposed to be the first intrinsic candidate of MTI \cite{lee2013crystal, rienks2019large,zhang2019topological,li2019intrinsic, otrokov2019prediction, aliev2019novel,gong2019experimental,lee2019spin, yan2019crystal,zeugner2019chemical,otrokov2019unique,deng2020quantum,liu2020robust,hao2019gapless,chen2019topological,li2019dirac,li2020competing,ding2020crystal}. It is composed of Te-Bi-Te-Mn-Te-Bi-Te septuple layers (SL) which are weakly bonded by van der Waals force as shown in Fig. \ref{cvt-xrd}(a). The van der Waals nature makes it possible to exfoliate a bulk MnBi$_2$Te$_4$ crystal into nm-thick thin flakes where QAHE was observed at a record high temperature in 5-layer devices \cite{deng2020quantum,liu2020robust}. Previously, MnBi$_2$Te$_4$ single crystals were grown via the flux method with Bi$_2$Te$_3$ \cite{yan2019crystal} or MnCl$_2$ \cite {deng2020quantum} as the flux or from the melt of the stoichiometric mixture \cite{otrokov2019prediction}. Other members of the MnBi$_{2n}$Te$_{3n+1}$ (MBT) family ($n>$1) can also be synthesized using the Bi$_2$Te$_3$ flux \cite{aliev2019novel,147,tian2019magnetic,1813, deng2020high-temperature,shi2019magnetic,ding2020crystal,wu2019natural,yan2020type,gordon2019strongly,hu2020universal,xu2019persistent,jo2020intrinsic,rienks2019large,tan2020metamagnetism,klimovskikh2020tunable,zhao2020phase,vidal2020orbital}. 

Chemical defects are observed in the MBT family. For flux-grown MnBi$_{2}$Te$_{4}$, studies find that 18(1) \% of Mn sites are occupied by Bi atoms \cite{ding2020crystal} while only 1-4\% of Bi sites are occupied by Mn atoms \cite{zeugner2019chemical,ding2020crystal,huang2020native,yuan2020electronic,yan2019evolution}. Mn and Bi sites, as well as the antisite defects are displayed in Fig. \ref{cvt-xrd} (a). This chemical complexity leads to electron carrier concentration on the order of $10^{20}$ cm$^{-3}$ in samples grown by flux method or from stoichiometric melting \cite{lee2019spin,yan2019crystal,gao2021layer}. Besides making the sample heavily-$n$ doped, chemical defects have profound impacts on the magnetism and band topology of MBT compounds. Antisites result in additional Mn sublattices in MBT. While the effect is relatively weak in MnBi$_2$Te$_4$, it is exaggerated in MnSb$_2$Te$_4$\cite{liu2021site,wimmer2021mn}, Sb-doped MnBi$_2$Te$_4$ \cite{yan2019evolution} and Sb-doped MnBi$_4$Te$_7$ \cite{Sb_147} where sizable amount of Mn$_{\rm{Sb}}$ antisites are introduced when Sb atoms are present in the lattice, leading to ferrimagnetic ground states. From the aspect of band topology, large amount of Mn$_{\rm{Sb}}$ antisites can be detrimental to non-trivial band topology in MBT \cite{liu2021site, Sb_147}. Because the QAHE is proposed theoretically on the ideal MnBi$_{2}$Te$_{4}$ structure, the defects which have caused the aforementioned complications may hinder the exploration of QAHE in MBT. Therefore, growth of single crystals that have a lower carrier concentration, fewer defects and higher magnetic homogeneity is highly desired for the exploration of various emergent phenomena and future applications of the material. 

In this work, we report a new single crystal growth route of MnBi$_2$Te$_4$ via chemical vapor transport (CVT) using I$_2$ as transport agent. We find that a small thermal gradient is sufficient to drive the CVT growth and allows a good control of growth. Through magnetic, transport and spectroscopic measurements, we show the carrier density is greatly reduced in the CVT-grown samples. Chern insulator state is observed in a 6-SL device with the highest reported mobility. All evidence points to a new promising growth route so that enhanced functionality of the devices can be made.

\section{Experimental methods}
Single crystals of MnBi$_2$Te$_4$ were grown using CVT method with I$_2$ as the transport agent. For comparison, the same crystals were also grown using Bi$_2$Te$_3$ flux. Mn, Bi and Te were mixed at a molar ratio of 12:176:276, loaded into a crucible and sealed in a quartz tube under one-third atmospheric pressure of Ar. The sample was heated to 900$^\circ$C for 5 hours, moved to another furnace where it slowly cooled from 595$^\circ$C to 585$^\circ$C in four days, and was kept at 585$^\circ$C for three days. MnBi$_2$Te$_4$ single crystal were obtained by centrifuging the ampule to separate the crystals from the Bi$_2$Te$_3$ flux. X-ray diffraction was collected using a PANalytical Empyrean diffractometer (Cu K-$\alpha$ radiation) on the (00$L$) surface to confirm the MnBi$_2$Te$_4$ phase. Wavelength-dispersive spectroscopy (WDS) was conducted on the bulk sample to determine the elemental ratio of single crystals. To address the sensitive chemical nature of 2D MnBi$_2$Te$_4$ flakes, all fabrication processes were completed in an argon environment without exposure to air, chemicals, or heat in a glovebox/evaporator assembly. Thin flakes of MnBi$_2$Te$_4$ were mechanically exfoliated from the bulk crystal onto the SiO$_2/$Si wafers using Scotch-tape. We then evaporated gold contacts by a pre-made stencil mask. Finally, BN and graphite flakes were transferred onto the MnBi$_2$Te$_4$ flake as the top gate.

Magnetization data were measured in a Quantum Design (QD) Magnetic Properties Measurement System (QD MPMS). Electric resistivity for both bulk samples and devices were measured using the six-probe configuration in a QD DynaCool Physical Properties Measurement System (DynaCool PPMS). Thin rectangular resistivity bars were cut from the bulk samples. The sign of $\rho_{xy}$ is chosen so that positive slope of $\rho_{xy}$(H) means dominating hole carriers. Longitudinal and Hall voltages were measured simultaneously using standard lock-in techniques. The gate voltages were applied by Keithley 2400 source meters.

Scanning tunneling microscopy (STM) data was acquired using a customized Unisoku USM1300 microscope. Single crystals were cleaved at room temperature in ultra-high vacuum (UHV) pressure of about $1\times10^{-10}$ Torr and immediately inserted into the STM chamber where they were kept at ~4.5 K during the measurements. Spectroscopic measurements were made using a standard lock-in technique with 915 Hz frequency and bias excitation as detailed in figure captions. STM tips used were home-made chemically-etched tungsten tips, annealed in UHV to bright orange color before being used for imaging.

ARPES measurements were carried out at the Advanced Light Source (ALS) endstation 7.0.2.1, and Stanford Synchrotron Research Laboratory (SSRL) beam-lines 5-2 and 5-4. Data was taken with photon energies of 26 eV with linear horizontal polarization. Samples were cleaved \emph{in situ} and measured under ultrahigh vacuum below 3 $\times 10^{-11}$ Torr at SSRL 5-4, 4 $\times 10^{-11}$ Torr at SSRL 5-2, and 2 $\times 10^{-11}$ Torr at ALS. Data was collected with Scienta R4000, DA30 L, and R4000 analyzers at SSRL 5-4, SSRL 5-2 and ALS, respectively.

\section{Experimental Results} 

\subsection{Growth optimization}

\begin{figure*}
\centering
\includegraphics[width=6.7in]{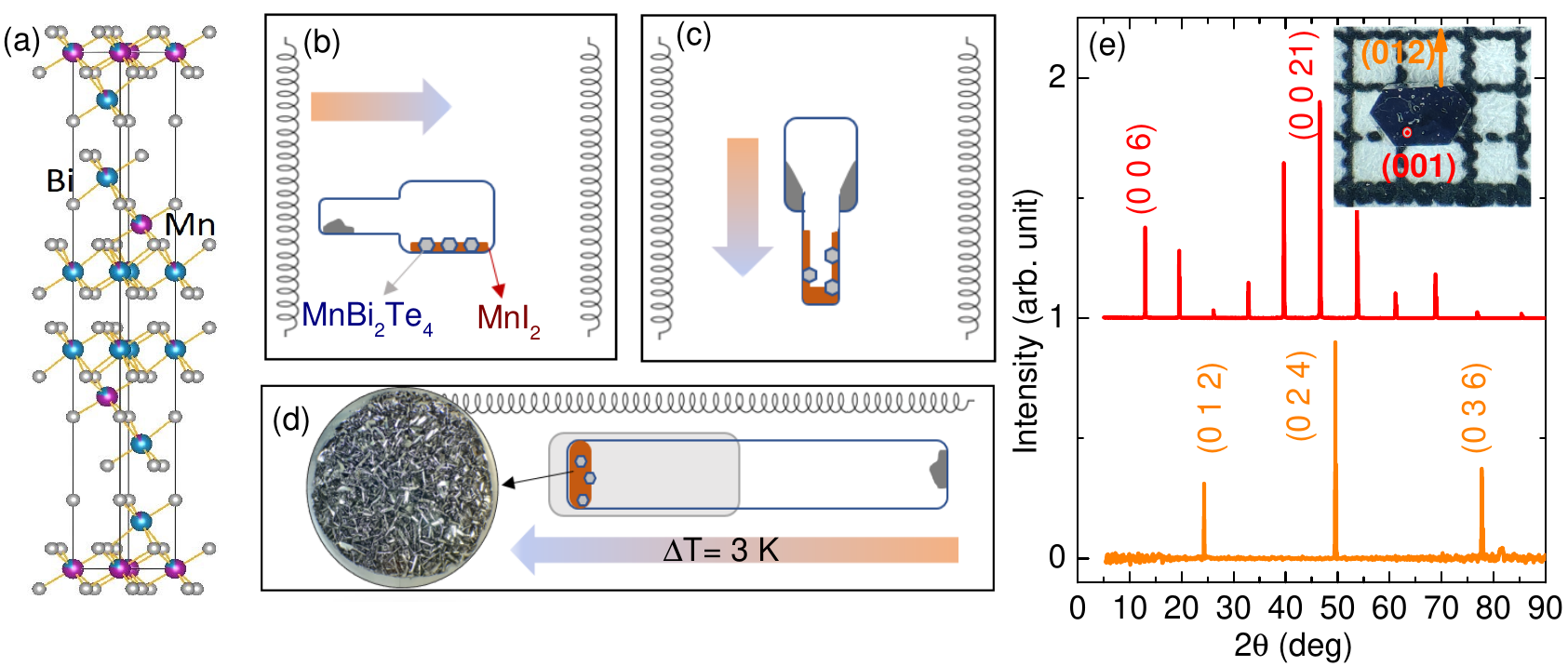} \caption{(a) Crystal structure of MnBi$_2$Te$_4$ with marked crystallographic Bi and Mn sites. Blue ball: Bi. Purple ball: Mn. Grey ball: Te. Mn site has Bi substitutions while the formation of Mn$_{\rm{Bi}}$ antisites results in a small amount of Mn atoms occupying the Bi site. The Mn atoms on the Mn site is denoted as Mn1, the Mn atoms on the Bi site is denoted as Mn2. Under $T_N$ Mn1 sublattice is AFM by itself; Mn2 sublattice is AFM by itself; Mn1 and Mn2 sublattices are AFM to each other. (b)-(c) Schematics for the CVT growth, using the internal horizontal and vertical temperature gradient in a box furnace, respectively. The arrow indicates the gradient direction. (d) Schematics of the growth in a fine-tuned three zone furnace. Inset: mm-lateral-sized single crystals with thickness from tens to hundreds of microns after being taken out of a 19-mm-diameter tube and rinsed. (e) X-ray diffraction spectrum on the (001) surface and the (012) surface of a CVT-grown single crystal. Inset: a hexagonal shaped single crystal from an one-week growth is shown on top of the 1$\times 1$ mm-grids. The as-grown surface orientations are indexed. }
\label{cvt-xrd}
\end{figure*}

Our initial trials of the CVT-growth of MnBi$_2$Te$_4$ were made in a Thermo Scientific muffle furnace rather than tube furnaces due to the two following considerations. First, the temperature profile of the CVT-growth is more delicate than the flux-growth of MnBi$_2$Te$_4$. The growth-end of the CVT-MnBi$_2$Te$_4$ should be kept at the temperature which is tested optimal for its flux growth. Second, a small temperature gradient of $\sim$ 2-3 K is sufficient and essential for the success of the CVT growth. As shown in Figs. \ref{cvt-xrd}(b) and (c), when the CVT-growths are positioned in the furnace, the thermal gradient intrinsic to the box furnace, either vertically between the top and bottom of the furnace, or horizontally between the heating element to the furnace center, is responsible and sufficient for driving the vapor transport. The temperature gradient is indicated by the arrows in both Figs. \ref{cvt-xrd}(b) and (c), which is only 2-3 degrees from source to end. 

Elemental form of Mn, Bi, Te and I are mixed according to the ratio of 1.7 : 2 : 4 : 1. Adding extra Mn or MnTe in the source end raises the melting point of the starting chunk, so it can remain as solid and allows a better control of its position at the source end during the transport. The elements were then sealed in a two-segment quartz tube as depicted in Fig. \ref{cvt-xrd}(b) or (c) under vacuum. The purpose of the two-segmented tube in a box furnace is to separate materials in the source (hot) and sink (cold) ends, and to have a balance over the cold-end area and transport rate so as to get sizable and abundant crystals. The quartz tube was then slowly heated to 900$^{o}$C overnight to avoid over-pressure. Afterwards, the tube is air-quenched and transferred back to this box furnace preset at 585$^{o}$C, which was pre-determined as the optimal temperature for our flux growth trials, and measured with an external thermocouple. The same thermocouple is used as the standard as we change between the furnaces, and as we measure the temperature gradient between the exact positions of hot and cold ends. In our trial-and-error process we find if the preset furnace temperature is 10 degrees higher, MnBi$_2$Te$_4$ in the cold end will be in the liquid form as condensed droplets; if the temperature is set to 10 degrees lower than the optimal, mixed phases of Bi$_2$Te$_3$ and MnBi$_{2n}$Te$_{3n+1}$ ($n\geq2$) will form. A short tube with a length of 8-10 cm was used to ensure the minimal temperature fluctuations. It is noted that all I$_2$ is reacted in the first step of slow warming so MnI$_2$ becomes the effective transport agent. After an optimal growth time of one to two weeks, mm-lateral-sized single crystals with thickness from tens to hundreds of microns are obtained at the cold end of the growth together with red MnI$_2$. Longer growth time yields larger crystals, but they are more likely to grow into each other. When samples are taken out, the mixtures from the cold end are rinsed with water to remove MnI$_2$ and to isolate the MnBi$_2$Te$_4$ crystals. A crystal grown with setup in Fig. \ref{cvt-xrd}(b) is shown on top of mm-grids in the inset of Fig. \ref{cvt-xrd}(e).

With the experience of growing MnBi$_2$Te$_4$ in the box furnace, we can also accommodate the growth in a three-zone tube furnace. Here, a very careful calibration is needed ahead of the time at the exact location of both ends of the growth ampule. The furnace is set so that the cold end has the same temperature as that in the mid-bottom of the box furnace in Figs. \ref{cvt-xrd}(b)(c) and the hot end is merely 3 K hotter. The cold end is then nested within an additional alumina crucible so that the temperature gradient can be further smoothed near the end as shown in Fig. \ref{cvt-xrd}(d). Eventually MnBi$_2$Te$_4$ and MnI$_2$ crystals form and almost cover the cold end of the tube. Figure \ref{cvt-xrd}(d) includes a picture showing a plate ``webbed" by MnBi$_2$Te$_4$ single crystals which were taken from a one-month growth in a 19-mm-diameter quartz tube and rinsed with water. With the careful calibration, the CVT-growth in the tube furnace can result in higher yields than both setups in box furnaces. The crystals in the ``web" can still be separated from each other for further measurements. 

Compared to the single crystal grown from other methods, the CVT samples tend to have a hexagonal shape with well-defined edges along the $a$ and $b$ directions. They can be up to 1 mm thick and appear more three-dimensional with flat and shiny edge surfaces. Hence in addition to the (00$L$) reflections, other reflections such as (012) can be observed on the as-grown surfaces on the side, as shown in Fig. \ref{cvt-xrd}(e). The two surfaces giving the XRD pattern are indexed for the crystal in the inset of Fig. \ref{cvt-xrd}(e). The WDS conducted over 15 pieces of hexagonal crystals from several CVT batches finds an elemental ratio of Mn$_{0.94(3)}$Bi$_{2.09(7)}$Te$_{4}$ with no significant batch-to-batch variation. There is piece-to-piece variation. We note that pieces with six well-defined edges as shown in the inset of Fig. \ref{cvt-xrd}(e) in general have higher Mn concentrations than those with one as-grown edge. The highest Mn ratio is up to 0.98. In comparison, the elemental analysis on our flux-grown crystals finds Mn$_{0.90(1)}$Bi$_{2.08(5)}$Te$_{4}$. This suggests an overall enhancement of Mn concentration, and motivates us to look into its impact on the physical properties.

\subsection{Physical properties}

\begin{figure}
\centering
\includegraphics[width=3.4in]{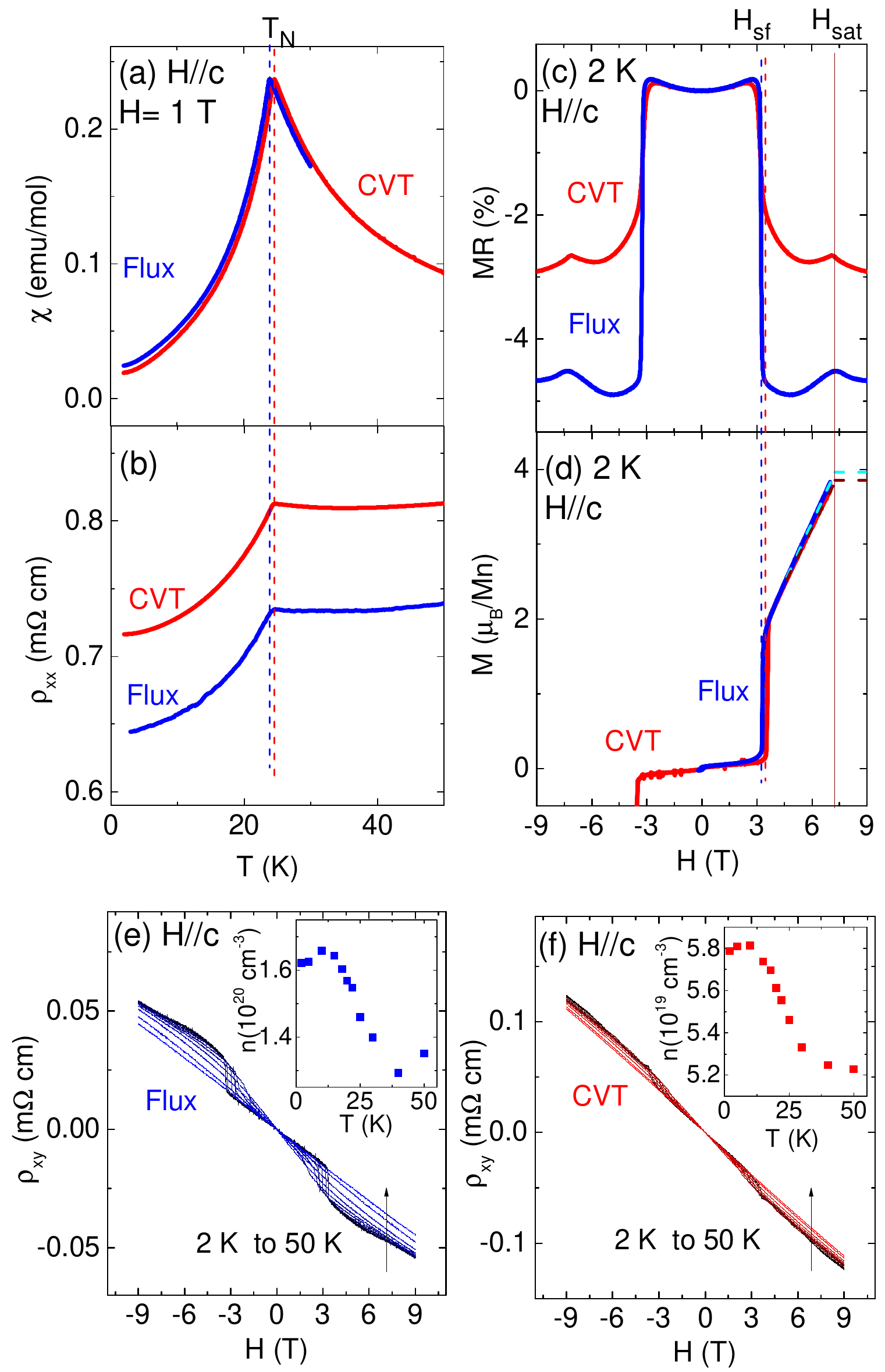} 
\caption{Comparison of the CVT-S1 and flux-S1 samples. (a)(b) Temperature-dependent magnetic susceptibility measured along H$_{//c}$ = 1 T and longitudinal resistivity $\rho_{xx}$ measured in the $ab$ plane. (c)-(d) Field-dependent MR and the magnetization with marked spin-flop transition field and saturation field measured at 2K. (e)-(f) Hall resistivity $\rho_{xy}$ measured from 2 K to 50 K for the two samples. Inset: Extracted carrier density from the slope of $\rho_{xy}(H)$ in the polarized FM phase at each temperature.}
\label{transport}
\end{figure}

The effect of higher Mn concentrations in CVT samples can be reflected in the bulk magnetic and transport measurements. The results in Fig. \ref{transport} are measured on and compared between a CVT-grown sample (CVT-S1) with Mn$_{0.95(1)}$Bi$_{2.09(1)}$Te$_{4}$ and a flux-grown sample (flux-S1) with Mn$_{0.90(1)}$Bi$_{2.11(2)}$Te$_{4}$, each characterized with WDS. Figures. \ref{transport} (a) and (b) show the temperature-dependent susceptibility $\chi_{H\|c}$ and resistivity $\rho_{xx}$ ($I//ab$). The AFM transition appears as a sharp kink in $\chi_{H\|c}$(T) in both panels, which is found to be 24.6 K for CVT-S1 and 23.8 K for flux-S1. The drop of $\rho_{xx}$ under $T_N$ is consistent with the reduced spin-disorder scattering due to the formation of the in-plane ferromagnetic (FM) order.
\begin{figure}
\centering
\includegraphics[width=3.5 in]{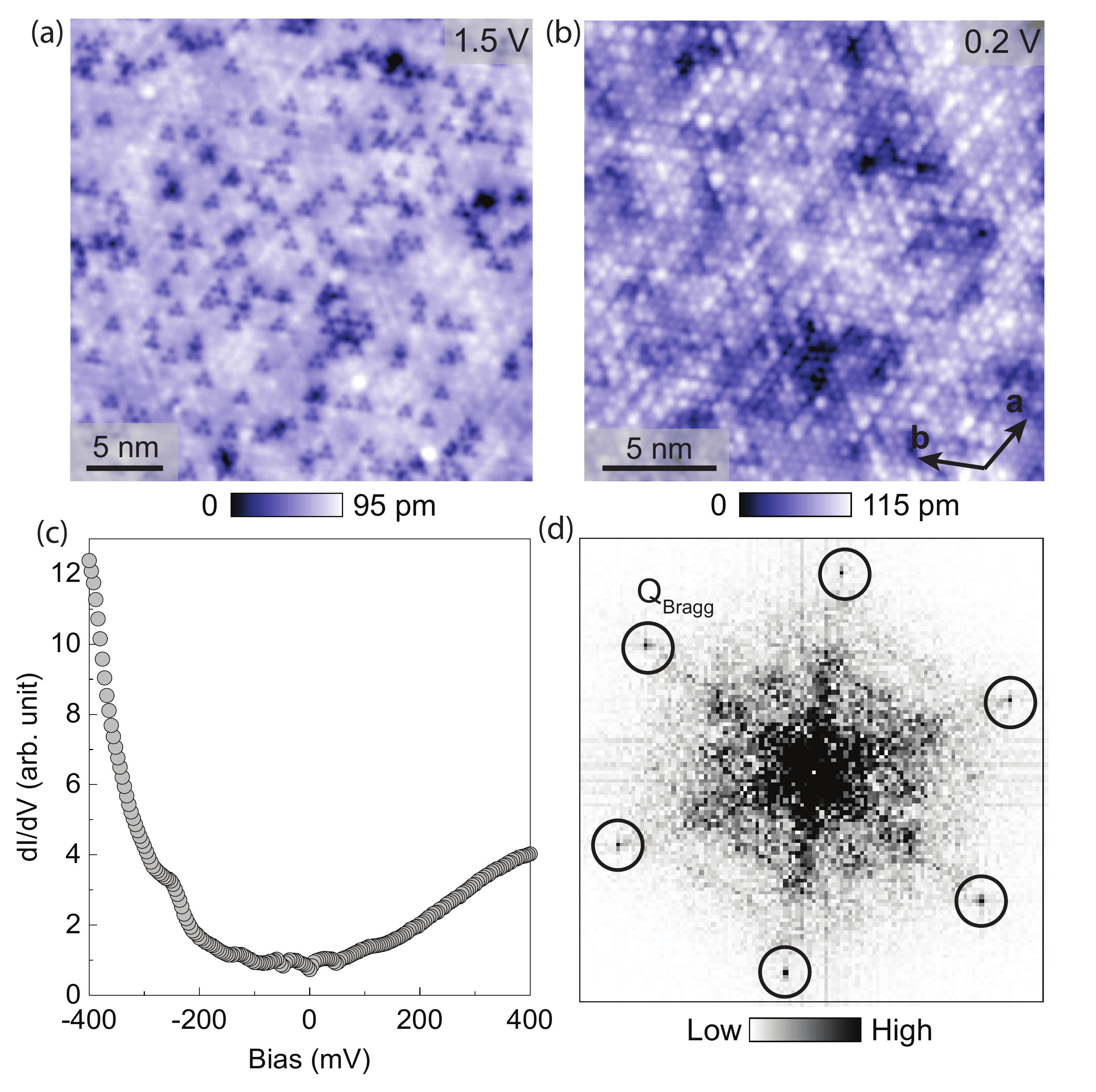} \caption{Scanning tunneling microscopy and spectroscopy of CVT-grown MnBi$_2$Te$_4$. (a) Large-scale STM topograph showing a large, flat surface obtained by the cleaving process.  Dark triangular features in the topograph represent Mn substitutions at the Bi site\cite{hor2010development}. (b) Zoom-in on a smaller region showing the expected hexagonal atomic structure. (c) Average dI/dV spectrum. (d) Fourier transform of the topograph in (b), with the atomic Bragg peaks denoted by black circles. STM setup condition: (a)1.5 V/1.5 nA ; (b) 200 mV/200 pA; (c) 400 mV/300 pA (4 mV bias excitation). All data is acquired at 4.5 K.}
\label{stm}
\end{figure}

The magnetism of MnBi$_2$Te$_4$ couples strongly with the charge carriers. The comparison of the magnetoresistance (MR), magnetization under field at 2K, as well as Hall resistivity from 2 to 50 K between the two samples is included in Figs. \ref{transport} (c)-(f). In Fig. \ref{transport}(d), the magnetization per Mn is calculated based on the Mn concentration obtained from WDS. At 3.3 T for flux-S1, a feature from the spin-flop transition shows up across all panels. The MR drops sharply with the sharp increase of the magnetic magnetization. In comparison, the CVT-S1 sample shows the same feature at 3.5 T. Then at 7.7 T, a feature due to the magnetic saturation is seen as a peak in MR and a subtle kink in Hall resistivity for both samples in Figs. \ref{transport} (c) (e) (f). Using the slope of Hall resistivity in the polarized FM phase, we calculate the electron-type carrier density at 2 K to be 5.79$\times10^{19}$ cm$^{-3}$ and 1.62$\times10^{20}$ cm$^{-3}$ for the CVT-S1 and flux-S1 sample at 2 K. The same calculation is carried out for each temperature. The temperature dependent carrier density is plotted in the insets of Fig. \ref{transport}(e) and (f). 

The magnetization data provides valuable insights on the distribution of Mn occupancy \cite{lai2021defect, Sb_147}. Bi substitutions of Mn on the Mn site as well as a very small amount of Mn$_{\rm{Bi}}$ antisites exist in MnBi$_2$Te$_4$. We denote the Mn atoms on the Mn site as Mn1, the Mn atoms on the Bi site as Mn2. Previous studies of the sample grown by the flux method \cite{lai2021defect} show that below 20 T, the Mn2 sublattice, aligns antiferromagnetically with the Mn1 sublattice. Therefore at $H_{sat}$, each Mn1 and Mn2 spins enter into its individual polarized FM state while these two sublattices are AFM to each other. Ref. \cite{lai2021defect} also shows at 50 T, Mn1 and Mn2 spins become parallel to each other. Therefore, the higher the Mn2-to-Mn1 ratio, the smaller the magnetic moment will be near 8 T since Mn1 and Mn2 are AFM to each other at 8 T. Considering the $M$(H) curve is linear between $H_{sf}$ and $H_{sat}$, we can estimate the moment at $H_{sat}$. The values are found to be 3.86 $\mu_B$/Mn for CVT-S1 and 3.94 $\mu_B$/Mn for flux-S1. We can then further quantitatively estimate the occupancy of Mn1 and Mn2 using

 \begin{figure}
\centering
\includegraphics[width=3.4in]{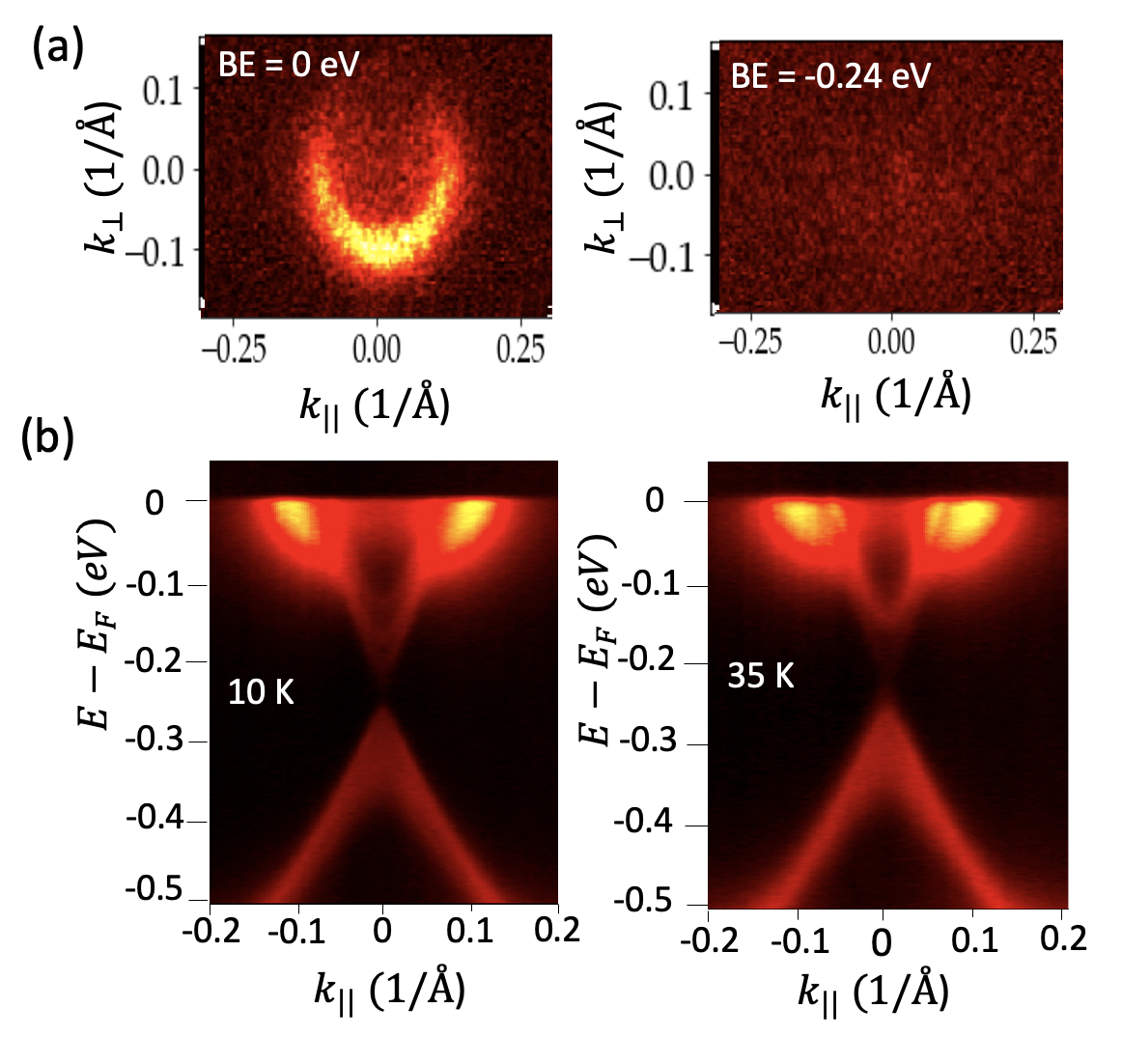} \caption{ARPES band maps and spectra on MnBi$_2$Te$_4$ sample CVT-S2. (a) ARPES intensity maps taken at E$_f$ and the Dirac point at -0.24 eV binding energy. (b) ARPES spectra taken at 10 K and 35 K on the $\Gamma$ to K cut, showing the TSS as well as a splitting of the bulk conduction band. ARPES data was taken with 26 eV, and the energy of the Dirac point is -0.24 eV, determined by finding the minimum of the energy distribution curve at $\Gamma$.}
\label{fig:ARPES}
\end{figure} 

\begin{equation}
    m_1+2m_2 = \text{Mn}_{\rm{WDS}}
\end{equation}
\begin{equation}
    \frac{m_1-2m_2}{{m_1+2m_2} } =\frac{M_{\rm{8T}}}{M_{\rm{50T}}} 
\end{equation}
Here $m_1$ and $m_2$ are the occupancy of Mn1 and Mn2, respectively. $\text{Mn}_{\rm{WDS}}$ is the Mn concentration obtained from WDS. $M_{\rm{8T}}$ is the initial magnetization value estimated from the $M-H$ curve at 8 T. $M_{\rm{50T}}$ is the final magnetization value when Mn2 are also polarized to align in parallel with Mn1, which is suggested to be $4.6 \mu_B$/Mn\cite{lai2021defect} based on the measurement up to 50 T. Based on these, $m_1$ and $m_2$ are found to be 0.835 and 0.032 for flux-S1, and 0.874 and 0.038 for CVT-S1. 

To observe the distribution of the defects and their effects on the electronic structure, STM was performed on a CVT-grown MnBi$_2$Te$_4$. The CVT-grown single crystals were cleaved at room temperature at the pressure of 1 $\times$ 10$^{-10}$ Torr and were immediately inserted into the STM head. Typical STM topographs show a flat surface with a hexagonal atomic structure (Fig. \ref{stm}(a,b,d)), consistent with the expected MnBi$_2$Te$_4$ topmost surface layer composed of Te atoms. Dark triangular features in the high-bias STM topograph in Fig. \ref{stm}(a) can be identified as Mn substitutions at the Bi site\cite{hor2010development}. By manual counting of individual defects observed in the topograph, we calculate the density of these substitutions in our CVT samples to be around 3.5\%. The antisite concentration is up from 3\% in the STM of the flux-grown sample \cite{yan2019crystal}, suggesting a similar trend we found from our magnetization data. Lattice constant extracted from the Fourier transform (FT) of the topograph (Fig. \ref{stm}(d)) is about 4.49 \AA. Average dI/dV spectra show a sharp upturn in conductance at around -200 mV in Fig. \ref{stm}(c). We note that this spectral feature is about 300 meV closer to the Fermi level compared to the spectra obtained in previous work\cite{yuan2020electronic,yan2019crystal}, which may indicate a lower level of self-doping in our samples.

The effect of defects and charge carriers on the band structure can be seen more clearly with ARPES. The measurements were made on three different CVT-samples. CVT-S2 has 6 as-grown edges while CVT-S3 and CVT-S4 have only one as-grown edge. The topological surface state (TSS), along with bulk conduction and valence bands were observed. Figure \ref{fig:ARPES} summarizes the ARPES data taken on CVT-S2. Figure \ref{fig:ARPES}(a) shows ARPES intensity maps at the Fermi energy, and at the Dirac point (DP) at -0.24 eV binding energy, respectively. Figure \ref{fig:ARPES}(b) shows the band structure of CVT-S2 cutting from the $\Gamma$ point to K point in the Brillouin zone above and below the N$\acute{\rm{e}}$el temperature near 24 K. A splitting of the bulk conduction band is clearly observed below the N$\acute{\rm{e}}$el temperature in the 26-eV spectra, similar to previous reports on samples grown by the flux method or stoichiometric melting \cite{chen2019topological,chulkov2020signatures}. The energies of the Dirac point can be estimated as the minimum of the energy distribution curve at the $\Gamma$ point. It is -0.24 eV for CVT-S2, -0.26 eV for CVT-S3, and -0.275 eV for CVT-S4. This variation in energy is consistent with what we have learned from our WDS measurements, that is, the well-shaped CVT crystals have higher Mn concentrations. In previous measurements on flux-grown samples, the DP energy ranges from -0.275 eV to -0.28 eV binding energy \cite{li2019dirac,hao2019gapless,kaminski2020gapless}, and samples grown with stoichiometric melting have DP energies at -0.27 and -0.275 eV binding energy \cite{chulkov2020signatures,chen2019topological}. Therefore, compared to flux-grown samples, the CVT-grown samples are in general more intrinsic. Especially, for the hexagonal-shaped CVT-2, the DP is around 35 meV lower and closer to the TSS. This is again consistent with the lower carrier concentration extracted from the Hall resistivity.

\subsection{Chern insulator state in 2D limit}

 \begin{figure}
\centering
\includegraphics[width=3.5in]{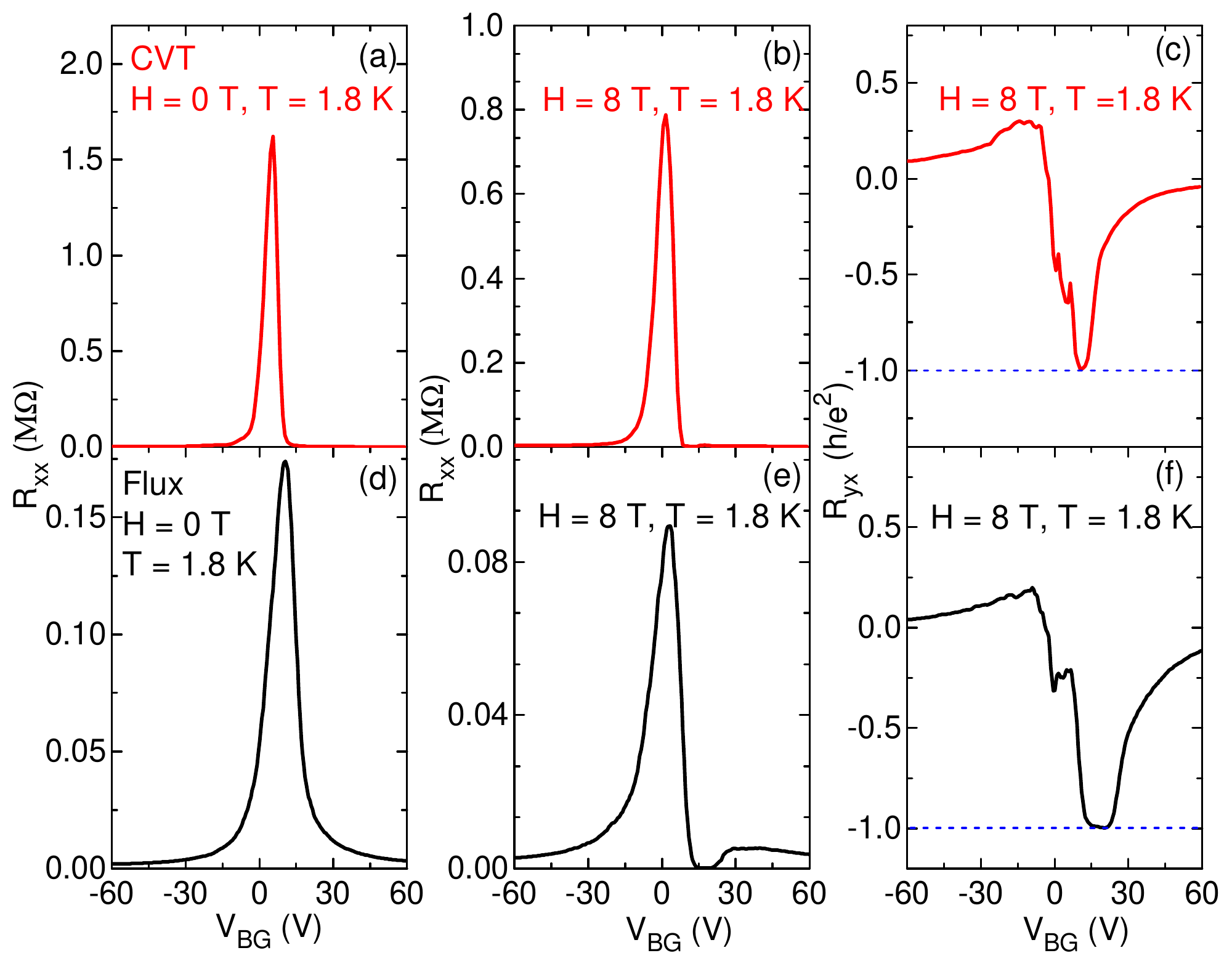} \caption{Observation of Chern insulator state in CVT and flux grown MnBi$_2$Te$_4$. All data are taken at 1.8 K. For the CVT-device and flux-device respectively: (a)(d) gate-voltage-dependent longitudinal resistance $R_{xx}$ in the AFM state at zero field. At the gated charge neutrality point, $R_{xx}$(T) in (a) is 10 times of that in (d) with a much sharper response to the gating voltage, suggesting better quality of the device. (b)(e) gate-voltage-dependent longitudinal resistance $R_{xx}$ in the FM state at 8 T, with marked reduction of resistance near the charge neutrality point. (c)(f) gate-voltage-dependent Hall resistance $R_{yx}$ in the FM state at 8 T, with marked quantized value of $-\frac{h}{e^2}$ near the charge neutrality point. }
\label{fig:QAH}
\end{figure}

To investigate the transport properties of the CVT-grown samples in the 2D limit, we exfoliated the MnBi$_2$Te$_4$ samples down to the atomically thin regime and fabricated 2D quantum devices. Figures \ref{fig:QAH} (a)-(c) show representative data of a 6-SL device made of a CVT grown MnBi$_2$Te$_4$ crystal. In the AFM phase ($H=0$), our transport measurement (Fig. \ref{fig:QAH} (a)) shows a clear insulating behavior with the resistance reaching over $10^{6}$ Ohms. This agrees with the theoretical expectation that the 6-SL AFM MnBi$_2$Te$_4$ is an Axion insulator with zero Chern number. Notably, as shown in Fig. \ref{fig:QAH} (a), the gate voltage that corresponds to the charge neutrality in the CVT-device is very close to $V_{BG}$=0 V. This indicates that natural doping in this 6-SL MnBi$_2$Te$_4$ device is small, consistent with the reduced carrier density that we obtained from the bulk transport measurements. Based on the field effect model, we estimate a carrier mobility of $\sim 2500$ cm$^2/$V$\cdot$s, which is the highest that has ever been reported. In the FM phase ($H$=8 T), our transport data reveal a fully vanishing longitudinal resistance $R_{xx}$ and fully quantized Hall resistance, which demonstrate the topological Chern insulator state, consistent with previous studies\cite{deng2020quantum,liu2020robust}.

\section{Discussion} 
WDS reveals higher amount of Mn concentration in CVT-S1 than that in flux-S1. 
Meanwhile from the analysis of the $M$(H) data, the concentrations of Mn1 and Mn2 sublattices, that is $m_1$ and $m_2$, are 0.835 and 0.032 for flux-S1, and 0.874 and 0.038 for CVT-S1, respectively. Ideally, MnBi$_2$Te$_4$ crystals with more Mn atoms going into the dominant Mn site (higher $m_1$) and less Mn atoms occupying the Bi site (lower $m_2$) will have better magnetic homogeneity and thus be more promising for the realization of QAHE, etc. However, our experiments indicate that the higher the total Mn concentrations we pushed in by optimizing the growth condition and method, the higher the $m_1$ and $m_2$ though $m_2$ increases much slower than $m_1$. Therefore, there is no way that MnBi$_2$Te$_4$ crystals with simultaneously higher Mn1 concentration and lower Mn$_{\rm{Bi}}$ can be made. If so, when screening samples for the device fabrication, will the samples with total higher Mn concentration be more ready to show quantized Hall conductance in devices? To answer this question, we compare the 6-SL device using the CVT-grown MnBi$_2$Te$_4$ crystal with a 6-SL device using a flux-grown MnBi$_2$Te$_4$ crystal. As shown in Fig. \ref{fig:QAH}, the two devices show qualitatively similar behaviors, i.e., an insulating behavior in the AFM phase and a topological Chern insulator behavior with quantized Hall response in the FM phase. 
However, one can clearly see that the resistance peak is much sharper and higher for the CVT-device as a result of the larger carrier mobility. Based on the field effect model, carrier mobility of the flux-device is found $\sim 800$ cm$^2/$V$\cdot$s, which is around one third of the CVT one. We further note that the Chern insulator behavior appears in a narrower range of gate voltage in the CVT-device. This may be related to the larger amount of antisite disorder present in the CVT-crystal. Despite this, the CVT-devices have a higher success rate in realizing the Chern insulator state with the same carefully-controlled fabrication process. We realized Chern insulator state in 6 out of 18 flux samples and 4 out of 5 CVT samples. The success rate of the device made from CVT-samples is more than twice as large as that of flux sample, which is likely to be attributed to the high Mn1 concentration that gives higher magnetic homogeneity. Therefore, the higher the $m_1$, the higher the success rate in realizing the Chern insulator state; the higher the $m_2$, the narrower the gate voltage window the Chern insulator state will appear.

The fact that both $m_1$ and $m_2$ are higher in CVT-S1 also explains why the magnetic transition temperature of CVT-S1 is higher. The higher Mn occupancy can also be associated with the reduced carrier
concentration in the CVT-S1, since Bi atoms on Mn site is an electron-donor and makes the sample more $n$-type, while Mn on Bi site is an electron-acceptor\cite{eisenbach2021fermilevels} which makes the sample more $p$-type. Compared to the flux-S1 sample, the Mn1 occupancy of the CVT-S1 sample increases by 0.039 while the Mn2 occupancy only increases by 0.006. This decreases the number of electron-donors and increases the number of electron-acceptors, thus making the sample less $n$-doped. Indeed, the low carrier concentration is universal in CVT single crystals. At 2 K, the carrier concentrations from 15 measured pieces from different batches range from 2.7-10 $\times10^{19}$ cm$^{-3}$ with an average of 5.8 $\times10^{19}$ cm$^{-3}$. The values from 5 measured pieces from different flux batch range in 1.3-2.0 $\times10^{20}$ cm$^{-3}$ with an average of 1.6$\times10^{20}$ cm$^{-3}$.

\section{Conclusion} 

In summary, compared with the flux-grown MnBi$_2$Te$_4$ single crystals, the CVT-grown ones have less Bi substitution on the Mn site and slightly more Mn$_{\rm{Bi}}$ antisites, leading to a smaller carrier density and reduced energy difference between the Fermi level and the Dirac point in bulk CVT-sample. Furthermore, when exfoliated into 6-SL device, the CVT samples show by far the highest mobility of 2500 cm$^2$V$\cdot$s with vanishing $R_{xx}$ and quantized $R_{xy}$ at 8 T. Therefore, our new growth design readily allowed us to achieve the Chern insulator state in devices with higher quality and high success rate. This paves a new way to optimize crystal growth of MnBi$_2$Te$_4$ and related MnBi$_{2n}$Te$_{3n+1}$ ($n\geq$2) to investigate the emergent phenomena arising from the interplay of topology and the magnetism. Future optimization such as varying the Mn concentration in the source end, using other transport agents such as TeI$_4$, etc. can be explored to further improve the sample quality.\\

 $\it{Note:}$ We noticed a similar vapor transport growth study on arXiv: 2110.06034 during the review process\cite{yan2021vapor}. The authors have made MnBi$_2$Te$_4$, Sb doped MnBi$_2$Te$_4$, MnSb$_2$Te$_4$, Mn$_{0.88(1)}$Bi$_{1.77(1)}$Sb$_{2.39(1)}$Te$_{6.96(1)}$ single crystals using I$_2$ as the transport agent. Other transport agents such as MnCl$_2$, TeCl$_4$, MnI$_2$ and MoCl$_5$ were explored. They also found that comparing to the flux grown samples, the samples grown by the vapor transport method tend to be Mn stoichiometric. They also suggest a small temperature gradient is necessary for the successful growths.

\section*{Acknowledgments}
 We thank Peipei Hao, Garrison Linn, Makoto Hashimoto, Donghui Lu, Chris Jozwiak, and Jonathan Denlinger for the help on ARPES measurements and useful discussions. Work at UCLA was supported by the U.S. Department of Energy (DOE), Office of Science, Office of Basic Energy Sciences under Award Number DE-SC0021117. Work at Harvard was supported by the Center for the Advancement of Topological Semimetals, an Energy Frontier Research Center funded by the U.S. Department of Energy (DOE) Office of Science, through the Ames Laboratory under contract DE-AC0207CH11358 (device fabrication) and by the STC Center for Integrated Quantum Materials (CIQM), NSF Grant No. DMR-1231319 (measurements). Work at CU-Boulder was funded by the U.S. DOE, Office of Science, Office of Basic Energy Sciences under Award Number DE-FG02-03ER46066. The ARPES work used resources of the Advanced Light Source, a U.S. DOE Office of Science User Facility under contract no. DE-AC02-05CH11231. Use of the Stanford Synchrotron Radiation Lightsource, SLAC National Accelerator Laboratory, is supported by the US DOE, Office of Science, Office of Basic Energy Sciences under Contract No. DE-AC02-76SF00515. Work at Boston College was supported by the National Science Foundation grant no. NSF-DMR-1654041 for the STM work. CH thanks the support by the Julian Schwinger Fellowship at UCLA.

\medskip

\bibliographystyle{apsrev4-1}
\bibliography{CVT124}
\end{document}